\documentclass[12pt,aps,prb,preprint]{revtex4}
\usepackage{graphicx}
\usepackage{dcolumn}
\usepackage{amsmath}
\usepackage{amssymb}
\usepackage{float}
\def\rv{{\bf r}}

\def\el{\ell_{osc}}

\def\beq{\begin{equation}}
\def\eeq{\end{equation}}

\begin{document}

\title{The Zel'dovich effect  in harmonically trapped, ultra-cold quantum gases}
\author{Aaron Farrell, Zachary MacDonald and Brandon P.~van Zyl} 
\affiliation{Department of Physics, St. Francis Xavier University, 
Antigonish, NS, Canada B2G 2W5} 
\begin{abstract}
We investigate the Zel'dovich effect in the context of ultra-cold, harmonically trapped quantum gases.   We suggest that currently available
experimental techniques in cold-atoms research offer an exciting opportunity for a direct observation of the Zel'dovich effect without the difficulties imposed by conventional
condensed matter and nuclear physics studies.
We also demonstrate an interesting scaling symmetry in the level rearragements which has heretofore gone unnoticed.
  \end{abstract}
\maketitle
\section{Introduction}

The Zel'dovich effect (ZE)~\cite{Zel} occurs in {\em any} quantum two-body system for which the constituent particles are under the influence of a long range attractive potential,
suplemented by a short-range attractive two-body interaction, which dominates at short distances.   The system first found to exhibit the ZE consists of an electron experiencing an 
attractive long-range Coulomb potential, which at short distances, is modified by a short-range interaction.~\cite{Zel}

The characterizing feature of the ZE in this scenario is that as the strength of the attractive two-body interaction reaches a critical value
({\it i.e.,} when a two-body bound state is supported in the short-range potential 
alone), the $S$-wave spectrum of the distorted Coulomb problem evolves such that the ground state $1S$ level
plunges down to large negative energies while simultaneously, the first radially excited $2S$ state rapidly falls to fill in the ``hole'' left by the ground state level.  
This also occurs for higher levels in which, generally, the $(n+1)S$ level replaces the $nS$ level.  This so-called level rearrangement {\em is} the signature of the ZE, and continues as the strength of 
the two-body interaction is further increased to support
additional low-energy scattering resonances (see e.g., Fig.~2 of Ref.~[\onlinecite{Zel}]).

Recently,  Combescure {\it et.~al} have revisited the ZE in the context of ``exotic atoms'',~\cite{comb1,comb2} where a negatively charged hadron replaces the electron, and the short-range interaction is 
provided by the strong nuclear force.
However, tuning the short-range interaction in exotic atoms implies that one must be able to adjust the nuclear force
in the laboratory, which is a formidable task.  Indeed, while it is theoretically easy to adjust the strength of the short-range interaction between the particles in any of the systems above,  
the experimental reality is very different.  As a result, direct experimental observation of the ZE has
been lacking, in spite of suggestions for its observation in quantum dots,~\cite{comb1} Rydberg atoms,~\cite{Kolo} and
atoms in strong magnetic fields.~\cite{karnakov}

In this article, we explore the possibility for a direct observation of the ZE in harmonically trapped, charge neutral, ultra-cold
atomic gases.  The charge neutrality of the atoms ensures that the long-range attractive potential is provided solely by the isotropic harmonic oscillator trap, 
while the short-range two-body interaction is naturally present owing to the two-body $s$-wave scattering, which is known to dominate at ultra-cold temperatures.  
Moreover, the short-range 
interaction between the atoms is completely tuneable in the laboratory {\it via} the Feshbach resonance.~\cite{cohen}  The multi-channel Feshbach resonance can be treated in a simpler 
single-channel model by a finite-range, attractive two-body interaction, supporting scattering resonances.  Thus  ultra-cold atoms, at least in principle, provide
all of the necessary ingredients for the experimental observation of the Zel'dovich effect.

The plan for the remainder of this paper is as follows. In Section II, we establish a deep connection between the level rearrangements and the two-body energy spectrum as
characterized by the $s$-wave scattering length, $a$.  This connection allows us to make contact with recent experimental results on ultra-cold two-body systems,~\cite{stoferle} from which we suggest that a direct
observation of the ZE is possible.  Then, in 
Section III, we investigate the influence of the range of the two-body interaction on the level rearrangements examined in Section II.  In particular, we reveal an interesting scaling symmetry 
of the two-body energy spectrum which has not been noticed before.
Finally, in Section IV, we present our concluding remarks. 
   
   \section{Zel'dovich Effect in Ultra-Cold Atoms}
 \subsection{Universal two-body energy spectrum} 
 The two-body spectrum for a pair of harmonically trapped ultra-cold atoms is obtained from the following Hamiltonian,
 \begin{equation}
 H = \frac{{\bf p}_1^2}{2M} +\frac{{\bf p}_2^2}{2M}+ \frac{1}{2}M\omega^2\rv_1^2+ \frac{1}{2}M\omega^2\rv_2^2+V_{SR}(|\rv_1-\rv_2])~,
 \end{equation}
where each atom has a mass of $M$ and $V_{SR}(|\rv_1-\rv_2|)$ is a short-range potential.  Introducing the usual relative, $\rv = \rv_1-\rv_2$, and centre of mass, ${\bf R} = (\rv_1+\rv_2)/2$ coordinates, and noting that the centre of mass motion may be separated out, the associated Schr\"odinger equation in the $s$-wave channel reads~\cite{farrell2}
 \begin{equation}
 -\frac{\hbar^2}{M} u''(r)+ \frac{1}{4}M\omega^2r^2u(r)+V_{SR}(r)u(r)+\frac{\hbar^2}{M}\frac{(d-1)(d-3)}{4r^2}u(r)=Eu(r)~,
 \end{equation}
 where $u(r)=r^{(d-1)/2}\psi(r)$ is the reduced radial two-body wave function, primes denote derivatives, $d$ is the dimension of the space, and $E$ is the relative energy of the two-body system. Defining the dimensionless variables $\eta =2E/\hbar\omega$, $\el = \sqrt{\hbar/M\omega}$ and $x= r/\sqrt{2}\el$, Eq.~(2) may be written as       
 \begin{equation}\label{TISE}
 -u''(x)+x^2u(x)+ \tilde{V}_{SR}(x)u(x)+\frac{(d-1)(d-3)}{4x^2}u(x)-\eta u(x)=0,
 \end{equation}
 where $\tilde{V}_{SR}(x) = 2V_{SR}/\hbar\omega$. 
  
Exact analytical solutions to (\ref{TISE}) exist $\forall d$ if the potential is taken to be an appropriately regularized zero-range contact interaction. For  $d=3$ the spectrum is described by\cite{busch, shea2008}
 \begin{equation}\label{otd}
 \frac{a}{\el} = \frac{\Gamma(1/4-E/(2\hbar\omega))}{\sqrt{2}\Gamma(3/4-E/(2\hbar\omega))},
 \end{equation} 
 for $d=1$ we have,~\cite{farrell2}
  \begin{equation}\label{1d}
 \frac{\el}{a} = \frac{\sqrt{2}\Gamma(3/4-E/(2\hbar\omega))}{\Gamma(1/4-E/(2\hbar\omega))},
 \end{equation} 
and for $d=2$,~\cite{farrell2}
  \begin{equation}\label{td}
  \tilde{\psi}(1/2-E/(2\hbar\omega)) = \ln{\frac{\el^2}{2a^2}}+2\ln{2}-2\gamma~.
  \end{equation}
In the above, $a$ is the $s$-wave scattering length in $V_{SR}$ alone, $\Gamma(\cdot)$ is the gamma function, $\tilde{\psi}(\cdot)$ is the digamma function and $\gamma=0.577215665...$ is the Euler 
constant.~\cite{handbook}  Note that in any dimension, the two-body spectrum is universal in the sense that the relative energy, $E$,
is determined entirely by the scattering length.   Thus, even for a two-body potential with {\em finite} range, $b$,  it has been shown that provided $b\ll\el$ (practically speaking, $b/\el \lesssim 0.01$), the {\em same} two-body energy 
spectrum as described above will be obtained for an arbitrary two-body interaction evaluated at the same scattering length.~\cite{farrell2}
\begin{figure}[ht]
\begin{center}
\includegraphics[scale=.3]{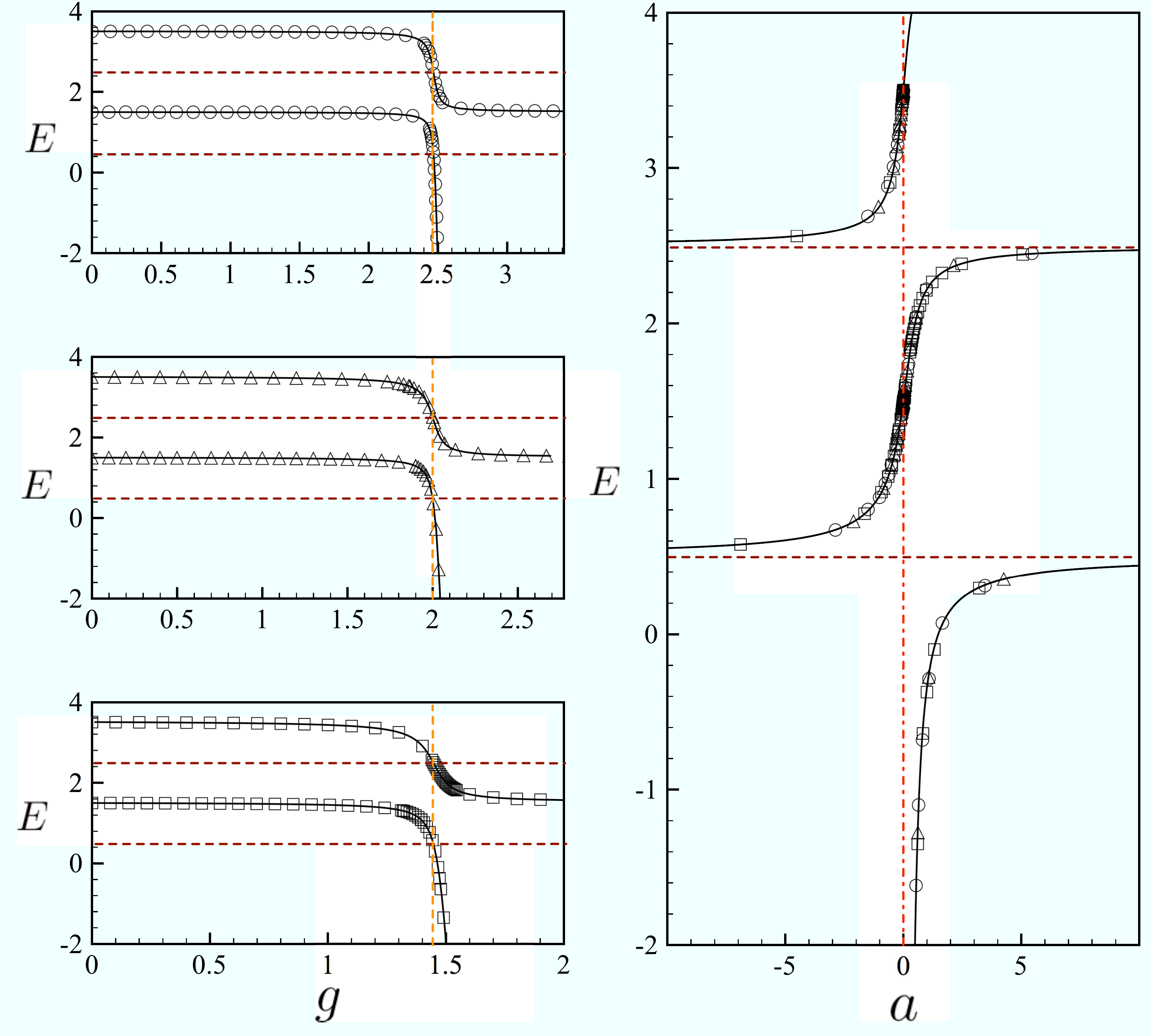}
\caption{Left panel: $s$-wave two-body energy spectrum versus strength for all 3 model potentials with fixed $b/\el=0.01$. Open circles, triangles and squares represent the numerical integration of Eq.~(3)
for the FSW, Poshl-Teller and exponential potentials, respectively. The vertical dashed line represents the critical strength, $g_c$, of all three potentials while the horizontal dashed lines are energy values at $a=\pm\infty$. The strength axis is scaled so that the critical strength values lie along the same vertical dashed line. Right panel: Energy versus scattering length for all 3 model potentials.  
The same symbols as the plots in the left panel are used. The vertical dot-dash line indicates
$a=0$. In all 4 plots, the solid black line is the exact expression obtained
from  
Eq. (\ref{otd}).  Units are scaled as discussed in the text.}
\end{center}
\end{figure}

\subsection{Level rearrangements}

In this section, the level rearrangements ({\it i.e.,} the ZE) exhibited by the two-body energy spectrum are investigated.  The results presented here are
strictly for three-dimensions (3D), although analogous findings are also observed in other dimensions.  We will focus on
three different interaction potentials, {\it viz.,}  a finite square well (FSW), the modified Poshl-Teller potential\cite{poschl} and an exponential potential\cite{exp}
\begin{equation}\label{srv}
 V_{SR}(r) =
\begin{cases}
-V_0\Theta(b-r)\\
-V_0\text{sech}^2{(r/b)}\\
-V_0\exp{(-r/b)}~,
\end{cases}
 \end{equation}
respectively.  In the above, $\Theta(\cdot)$ is the Heaviside step function, and $V_0$ is the depth of the potential. 
The 3D $s$-wave scattering lengths for the potentials are given by
(in the same order as the potentials listed above) \cite{shea2008, CJP,poschl, Flugge, exp, ahmed}
 \begin{equation}\label{scatlength}
  a=
\begin{cases}
b\left(1-\frac{\tan{\sqrt{g}}}{\sqrt{g}}\right)\\
b\left(\gamma + \tilde{\psi}(\lambda)+\frac{\pi}{2}\cot{\pi\lambda/2}\right)\\
b\left(2\gamma + \ln{g}-\frac{\pi Y_0(2\sqrt{g})}{J_0(2\sqrt{g})}\right)~,
\end{cases}
 \end{equation}
where $g\equiv MV_0b^2/\hbar^2$ is the dimensionless strength of the potential, $\lambda \equiv (1-\sqrt{1+4g})/2$, and $J_0(\cdot)$ and $Y_0(\cdot)$ are the zeroth order Bessel functions of the first and second kind, respectively.~\cite{handbook}

We proceed by  numerically integrating Eq.~(\ref{TISE}) for each of the  three potentials.  For our numerics, we have set $\hbar=\omega=1$ and $M=2$ to be consistent with the numerical results in 
Refs.~[\onlinecite{busch},\onlinecite{shea2008}].  
We plot our numerical results (open symbols) for the relative energy, $E$ (in units of $\hbar\omega$), as characterized by both the strength, $g$, and
the $s$-wave scattering length, $a$, in Figure
1.   The left panel illustrates the level rearrangements as the strength, $g$, is increased beyond the first scattering resonance, whereas the right panel illustrates the relative energy, $E$, as determined
by the $s$-wave scattering length.  The level rearrangements shown in the
left panels illustrate the $2S$ level replacing the $1S$ level at the first scattering resonance, while the $1S$ level dives down to large negative values.

A further examination of  Fig.~1 reveals that while $b/\el = 0.01$  for all three potentials,  the level rearrangements displayed in the 
left panels exhibit noticeable differences.  
In particular, we see that the FSW has a much sharper drop at $g=g_c$, than the Poshl-Teller or exponential potentials.  These level repulsions, or ``anticrossings'', are known to be as a 
result of the levels belonging to the
same $SO(2)$ symmetry of the Hamiltonian, while the mixing of the levels is dependent on how rapidly the short-range potential ``shuts off''.~\cite{comb1}

The underlying message here is as follows.  While all three plots on the left of Fig.~1 display the Zel'dovich effect, namely they all undergo level rearrangement at some value of the strength parameter $g$, all three {\em different} potentials map on to the same $E$ vs.~$a$ curve, as illustrated in the right panel of Figure 1. This reaffirms that while the details of the ZE
are sensitive to the form of the two-body interaction, the energy dependence on the scattering length, $a$, is indeed universal.
It is also worthwhile pointing out that the solid curves in the left panels of Fig.~1 are obtained from substituting the expressions for the scattering length, Eq.~(8), into the Eq.~(4), which is
exact {\em only} for a zero-range interaction.  However, it is clear that the numerically obtained open symbols closely follow the solid curve derived from Equation (4).
Thus,  for $b/\el \ll 1$, the level rearrangements in harmonically trapped two-body systems interacting {\it via} a finite, short-range potential, are all equivalent to a zero-range interaction.  Viewed another way,
given a set of data for
$E$ vs.~$a$, there must exist {\em some} quantum two-body system ({\it i.e.,} the two-body potential need not be known explicitly) whose $E$ vs.~$g$ dependence exhibits the Zel'dovich effect.  
This observation has some interesting implications,
which we further explore in the next subsection.            
     
\subsection{Flow of the Spectrum}

In order to make the connection between the $E$ vs.~$g$ and $E$ vs.~$a$ curves more apparent, we now study the ``flow'' of the two-body energy spectrum.   Although we focus our attention
to the FSW, the same analysis holds for any other potential. 
\begin{figure}[h]
\begin{center}$
\begin{array}{cc}
\includegraphics[scale=.35]{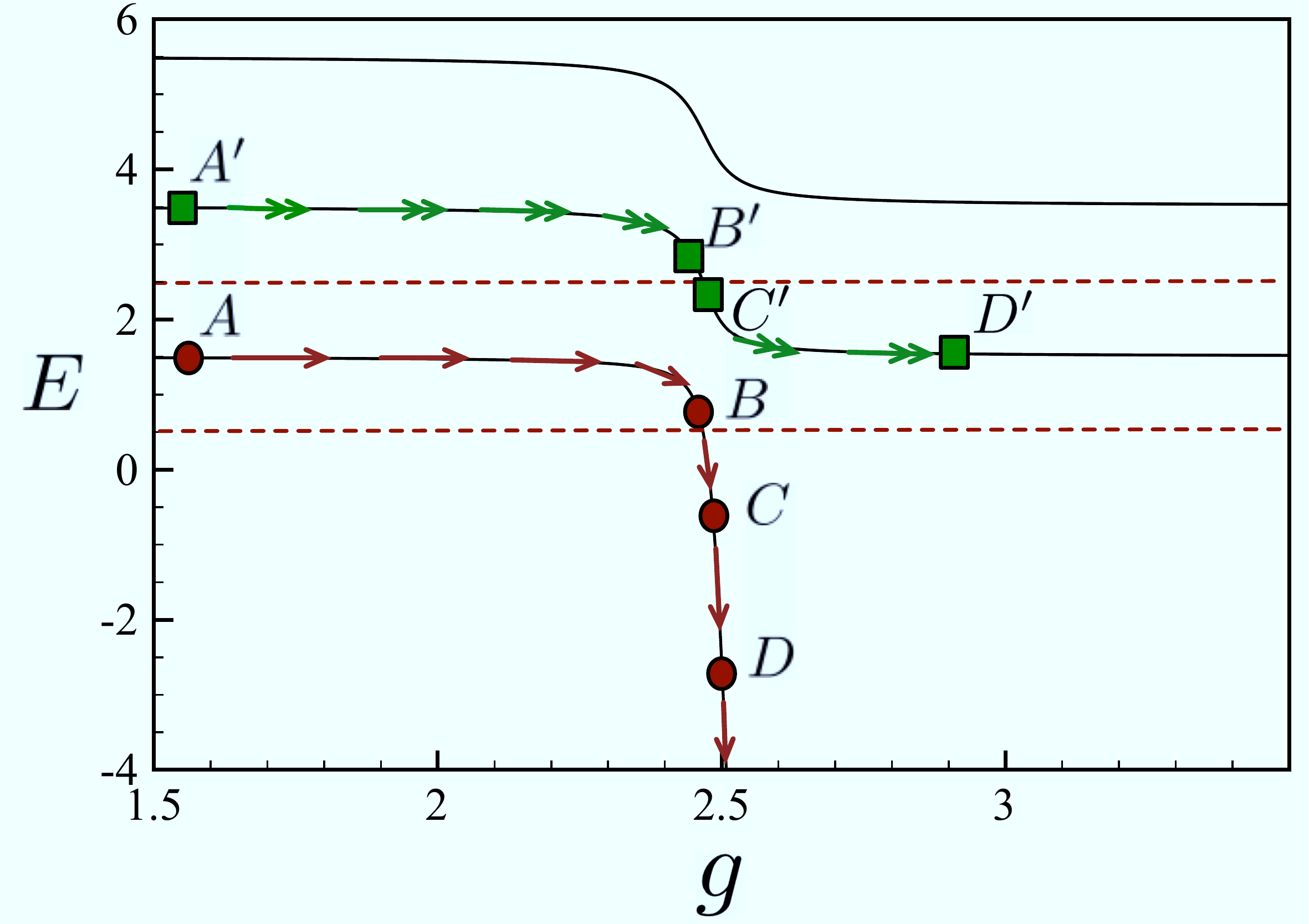}&
\includegraphics[scale=.35]{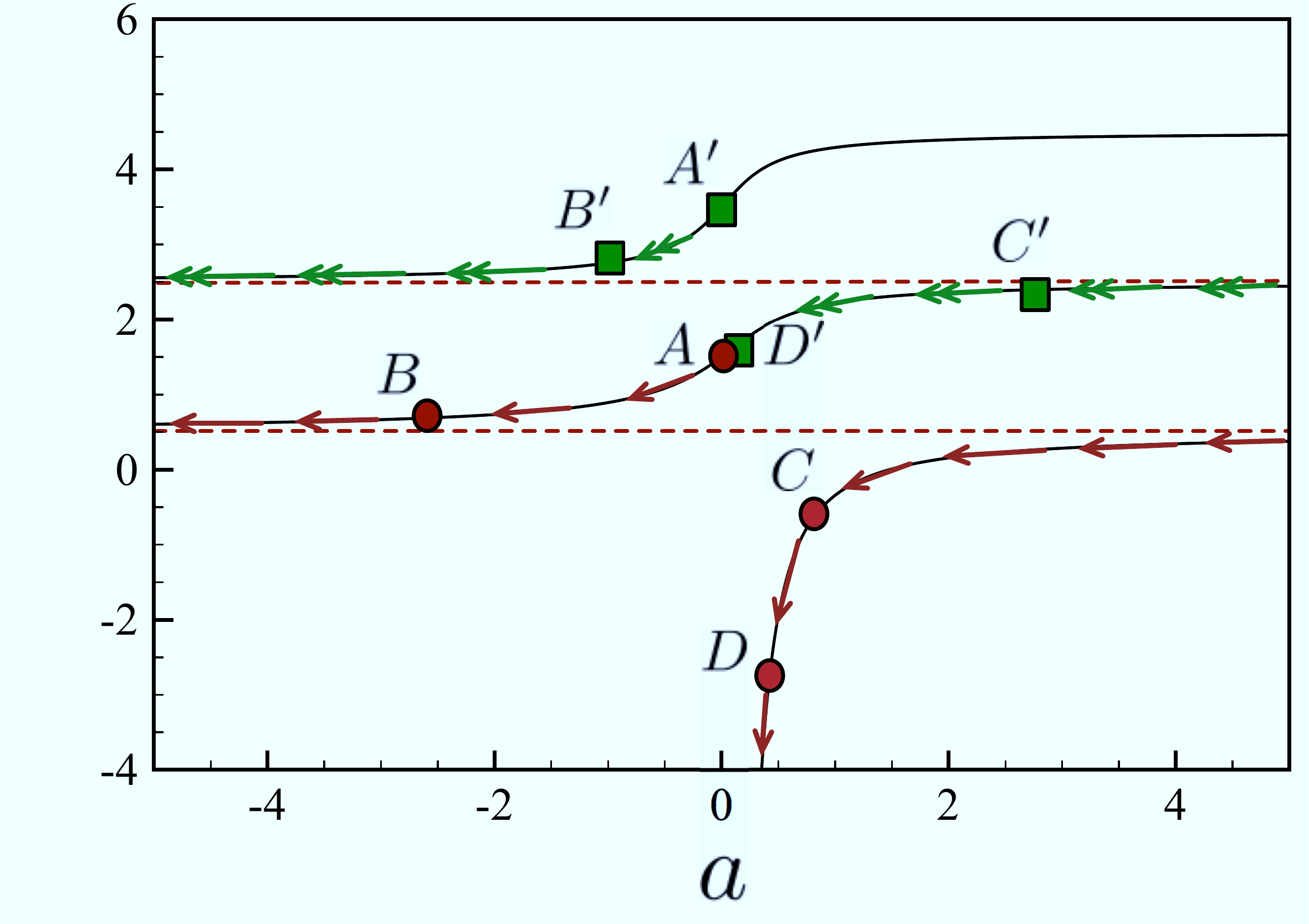}  \end{array}$
\end{center}
\caption{(Color online)  Flow of the spectrum for the FSW. Boxed and double arrows (green online) represent the flow of the first excited state while circles and single arrows (red online) follow the flow of the ground state. Identical points in each panel are labeled by the same letter. For example, the point $B$ on the left is the {\em exact same} data point as $B$ on the right but subject to the transformation in Equation (\ref{scatlength}).  The horizontal dotted lines in both panels correspond to the asymptotic values for the energy $E$ at $|a| = \infty$, as in Figure 1.  See also the discussion preceding Eq.~(19) in Section III.}
\end{figure}

In the left panel of Fig.~2, we note that as $g$ is increased from zero,  the energy only slightly varies from the unperturbed energy, until the critical strength, $g_c$, is reached at which point the Zel'dovich effect occurs.  In the right panel of Fig.~2, the same flow is illustrated, but this time in terms of the  scattering length.  The lower flow in the left panel (red online)  illustrates that the trajectory of the
ground state $A \to B \to C \to D$ is continuous though the resonance at $g=g_c$.  However, as we follow the same path in $E$ vs.~$a$, the point $B$ flows out to $a \to -\infty$ while $C$ and $D$ flow in
from $a \to +\infty$ and then to
$a \to 0$.  Thus, while the flow for the energy spectrum in $g$-space is continuous, the flow in $a$-space appears to be disconnected.  Similarly for the first excited state (green online) where the
$B'$ and $C'$ flow is continuous in $g$-space, but rapidly branches off to $a \to -\infty$ and $a \to 0$, in $a$-space, respectively.  The continuous flow in $g$-space suggests that the $a$-space spectrum is more appropriately viewed on the
topology of a cylinder, where $a=\pm \infty$ may be identified.

\subsubsection{Cylindrical Mapping}
The observations made above suggest that we map the $E$ vs.~$a$ spectrum onto the surface of a cylinder.  The details of this mapping are closely related to the mapping of
the real line (in our case, the scattering length) onto the unit circle, $S^1$, followed by constructing the Cartesian product, $\mathbb{R}\times S^1$, with $\mathbb{R}$ identified with the energy, $E$.  The essential point of this mapping is to provide a more natural interpretation for the two-body
$E$ vs.~$a$ spectrum. 

To this end,  Fig.~3 illustrates a series of  ``snapshots'' which show how the the original $E$ vs.~$a$ spectrum is mapped onto the surface of a cylinder.
\begin{figure}[h]
\begin{center}
\includegraphics[scale=.45]{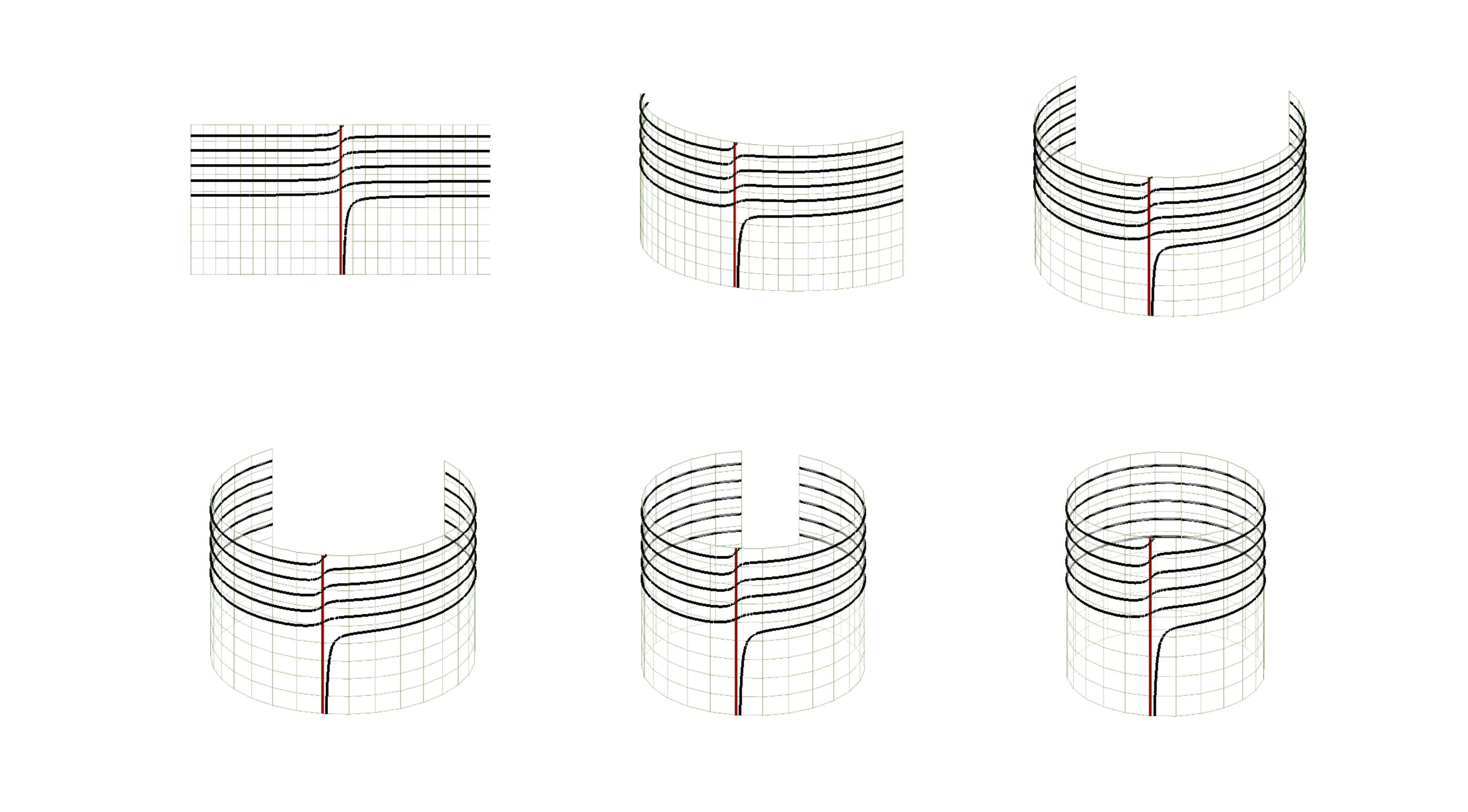}
 
  \caption{The $E$ vs.~$a$ spectrum being rolled onto a cylinder. Top from left to right:   $a=-10$ to $a=10$, $a=-30$ to $a=30$, $a=-35$ to $a=35$. Bottom from left to right: $a=-40$ to $a=40$, $a=-90$ to $a=90$, $a=-\infty$ to $a=\infty$. The thick vertical line (red online) represents $a=0$.  The symmtery axis of the cylinder is the $E$ axis
while the azimuthal angle is connected to the scattering length, $a$.}
\end{center}
 
 \end{figure}
Each of the 6 panels in Fig.~3 should be viewed as an intermediate step in taking the $E$ vs.~$a$ spectrum and rolling it onto a cylinder.    
In Fig.~4, we present the complete mapping of the $E$ vs.~$a$ spectrum up to the $4$-th excited state of the bare harmonic trap.  
This ``Zel'dovich spiral'' (ZS) may now be explicitly connected to the level rearrangements
discussed in the left panel of Fig.~1 above.  

Indeed, we observe that 
the flow of the spectrum shown in the left panels of Fig.~2 correspond to clockwise (CW) rotations about the Zel'dovich spiral. 
That is, increasing the strength of the two-body interaction corresponds to moving along the ZS in a CW direction, with the
starting point ({\it i.e.,} the front of the cylinder) along the thick vertical line (red online) in Figure 4. 
\begin{figure}[h]
\begin{center}
\includegraphics[scale=.5]{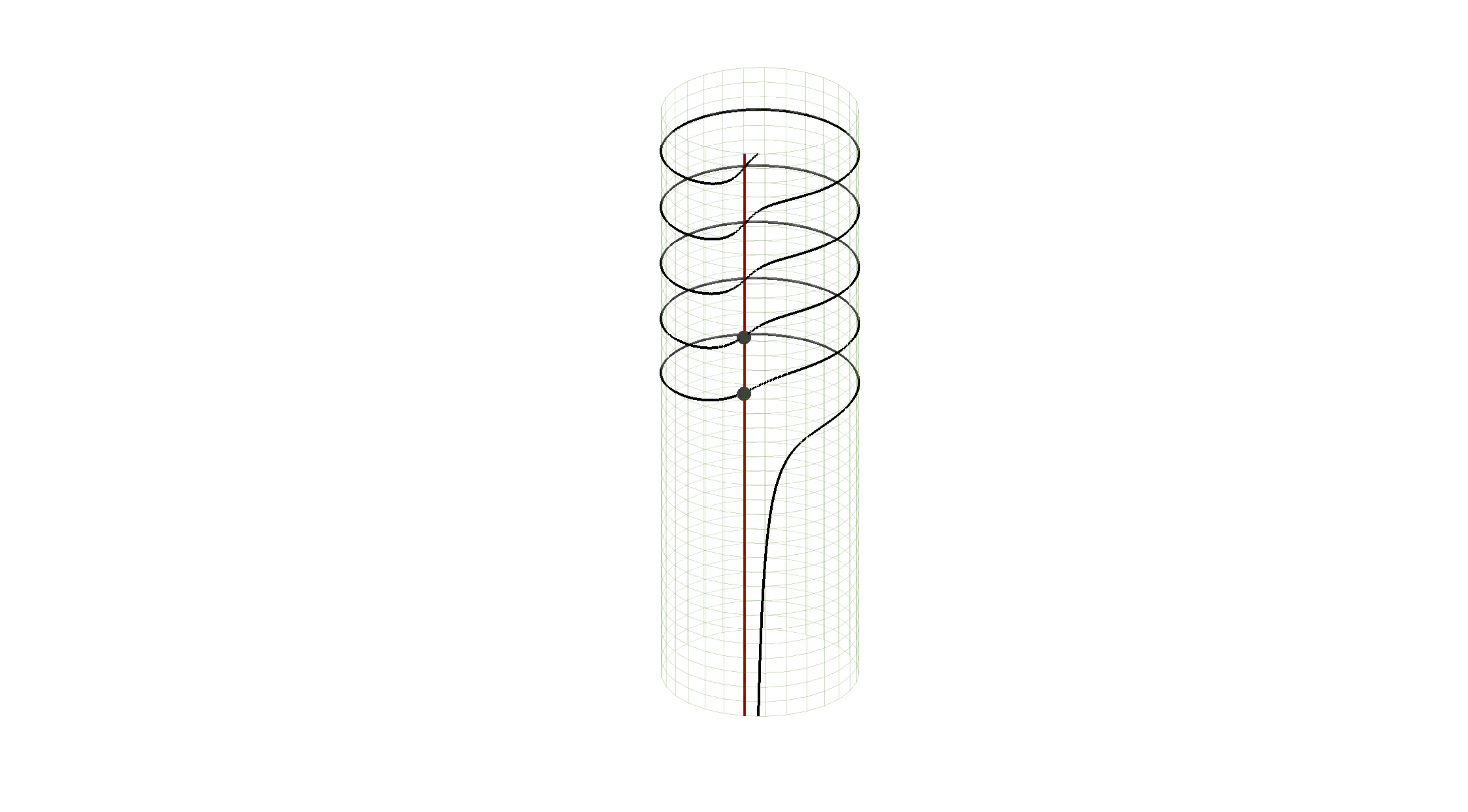}
 \caption{The complete mapping of the $E$ vs.~$a$ spectrum onto the surface of a cylinder illustrating the ``Zel'dovich spiral''. 
  The solid vertical line (red online) identifies the unperturbed system for which $a=0$.
 Every $2\pi$ winding along the spiral corresponds to a complete level rearrangement {\it e.g.,} $2S \to 1S$ after $2\pi$ rotations.  
The lower solid circle indicates the unperturbed $1S$ level, whereas the upper solid circle corresponds to the $2S$ level.}
\end{center}
 \end{figure}
A CW rotation of $\pi$ puts us on the back of the cylinder, or $a=-\infty$, whereas a counter-CW rotation of $\pi$ takes us to $a=+\infty$ ({\it i.e.,} the azimuthal angle $|\phi| = \pi$ is a branch point).  

To see how the ZS naturally contains the level rearrangements, let us first begin at $E=3/2$, which in Fig.~4 is represented by the lower solid circle along the vertical line.
As we move in a CW rotation along the spiral, the $E=3/2$ ($a=0$, $\phi=0$) goes to $E\to 1/2^{+}$ for large negative values of $a$, and finally to $E=1/2$ at  $a=-\infty$ (lower dotted curve in the right panel
of Figure 2).  
A further infinitesimal rotation takes us to $E=1/2^{-}$ at large positive values of $a$, and finally
to $E \to -\infty$ at $a = 0$ after a full $2\pi$ rotation; we have just followed the flow of the $1S$ level in the left panel of Fig.~1, {\it viz.,} $A \to B \to C \to D\to \cdot \cdot \cdot$.
Similarly,  the upper solid circle in Fig.~4 corresponds to $E=7/2$, which as we rotate CW, evolves to $E\to5/2^{+}$ for large negative values of $a$, $E=5/2$ at $|a|=\infty$ (upper dotted curve in the right
panel of Figure 2), and subsequently to $E=3/2$ at $\phi=2\pi$; this description
is precisely the flow of the $2S$ level in the left panel of Figure 2.  If we were to the continue with our CW rotation ({\it i.e.,} continue increasing the strength, $g$), we would then evolve from $E=3/2 \to 1/2$ at 
$\phi = 3 \pi$ followed by
$E \to -\infty$ at $\phi = 4 \pi$. 

In our opinion, viewing level rearrangements in this way is more natural than the original $E$ vs.~$a$ spectrum in $\mathbb{R}^2$.  
We see that critical strengths, $g_c$, correspond to CW rotations of odd multiples
of $\pi$ whereas a complete level rearrangement occurs for even multiples of $\pi$.  In general, the $(n+1)S$ level, with $E_n = 2n+3/2$ ($n=0,1,2,...$) will eventually evolve to the $1S$ level after $2n\pi$ CW
rotations along the spiral.   

The Zel'dovich spiral also helps to clarify several misconceptions about $E$ vs.~$a$ spectrum in the literature. The spectrum is typically understood by taking $a=\pm \infty$ separately, and assigning different interpretations to $a\to0^+$ and $a\to0^-$.  An example of this is a recent contribution is by Shea {\it et al},~\cite{shea2008} where the authors describe the spectrum by first ``starting from the far left" and making the interaction weaker and weaker as $a\to0^-$ and then {\em independently} ``starting from the right" and making the interaction stronger and stronger as $a\to0^+$.   
On the ZS, nothing is ambiguous, since one always moves in a CW rotation along the spiral, corresponding to increasing the strength, $g$, of the interaction; a complete level rearrangement
occurs after we undergo an even muliple of $\pi$ CW rotations.  Furthermore, the ``counter-intuitive'' properties of the $E$ vs.~$a$ spectrum discussed in Ref.~[\onlinecite{busch}] are now seen to be 
nothing more than a manifestation of the onset of the Zel'dovich effect.
We find it rather surprising that the ZE has been present in the two-body $E$ vs.~$a$ spectrum all along, but until now, has gone unnoticed.

\subsubsection{Experimental Observations}

In a recent work, St\"oferle {\it et. al},~\cite{stoferle} have experimentally measured the binding energy as a function of the $s$-wave scattering length between two interacting particles in a harmonic trap.  
This experiment highlights the versatility of trapped, ultra-cold atomic systems, in which an analytically solvable model, once only the purview of theoretical physics,
has now been realized in the laboratory.  Remarkably, the experimental 
results for the $E$ vs.~$a$ spectrum are in
excellent agreement  with theory, (see Fig.~2 in Ref.~[\onlinecite{stoferle}]), even though the two-body interaction in the experiments is most certainly {\em not} a zero range interaction.  
Thus, the theoretical prediction that the $E$ vs.~$a$ spectrum is universal has been confirmed experimentally.

What has not been appreciated until now, however, is that the experimental $E$ vs.~$a$ spectrum obtained in Ref.~[\onlinecite{stoferle}] is exactly equivalent to obtaining the ground state branch in the 
left panel of Fig.~2 (single arrows, red online).  In other words, the work of St\"oferle {\it et. al},  has already been a direct {\em experimental observation} of the ground state branch of the two-body system
exhibiting the Zel'dovich effect. We therefore suggest that further experiments along the lines of Ref.~[\onlinecite{stoferle}]  be performed so that data corresponding to the double arrows and primed letters
(green branch online) in the right panel of Fig.~2 may be obtained.  
If such an extension to the experiments in Ref.~[\onlinecite{stoferle}]  is viable, then according to our analysis, this data would be exactly equivalent to the $2S$ branch
(double arrows, green online, in the left panel of Fig.~2) undergoing the Zel'dovich effect.  Therefore, just a few additional data points in the $E$ vs.~$a$ spectrum, would provide for a direct experimental
confirmation of the ZE for two interacting particles confined in a harmonic trap.

\section{Level rearrangements in the zero-range limit}

We close this work with a discussion of an interesting scaling symmetry present in the  the level rearrangements.  Specifically, we show that in the $b \to 0$ limit, the {\em entire} two-body $E$ vs.~$g$ spectrum
is determined by only the first level rearrangement.   

Figure 5 illustrates the spectrum through the first three low-energy scattering resonances, {\it viz.,} $g=g_0, g_1, g_2$ for the FSW.
\begin{figure}[h]
\begin{center}
\includegraphics[scale=.25]{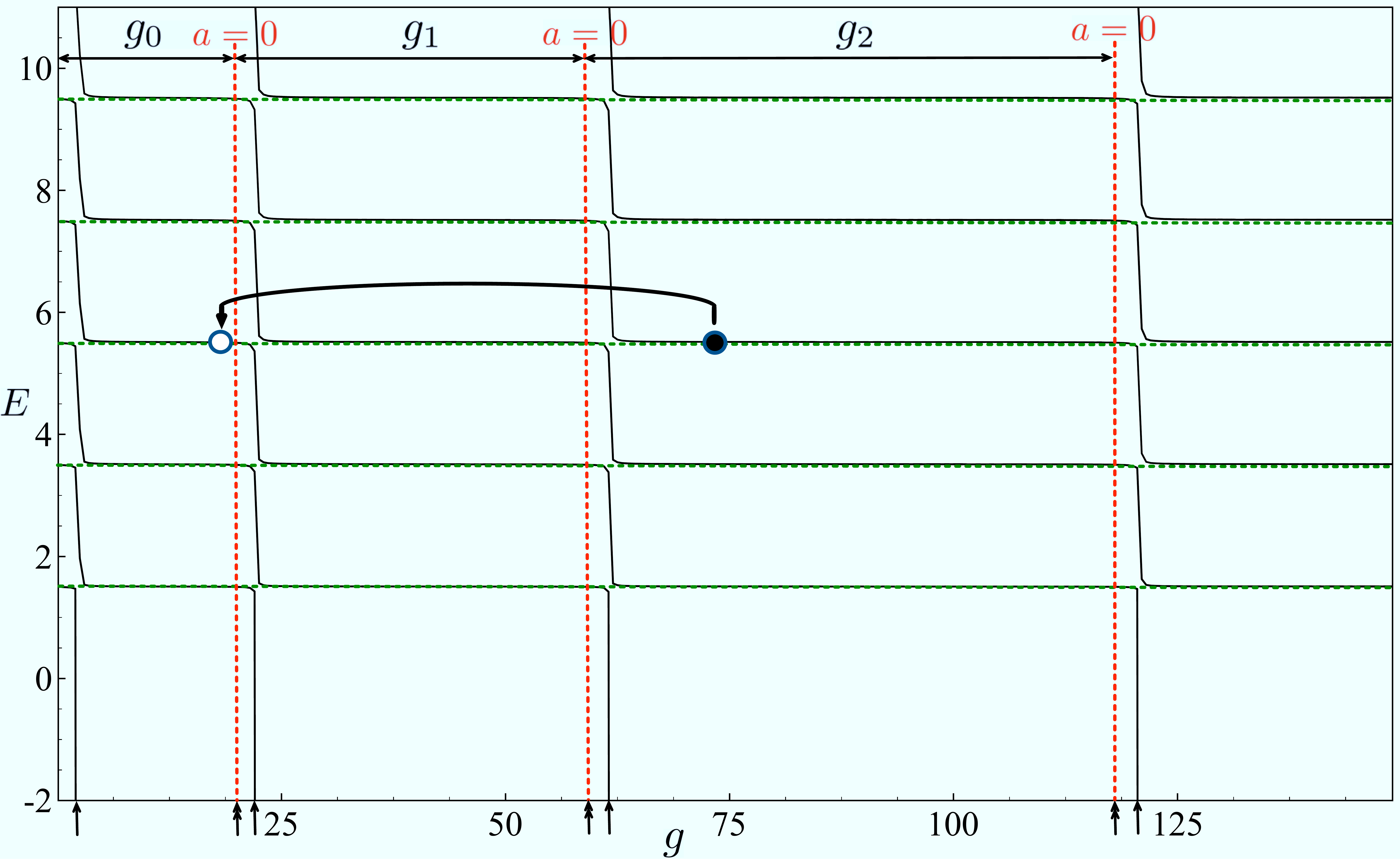}
 \caption{Several level rearrangements for a FSW interaction with $b/\el = 0.01$. Dotted horizontal lines  (green online) are unperturbed energy values while dotted vertical lines (red online) are values at which $a=0$.  Each subsequent region corresponds to a CW rotation of $2\pi$ along the Zel'dovich spiral.  The solid black circle is a representative data point which we wish to map back into the $g_0$ region,
as schematically illustrated by the open circle in $g_0$.  Single arrows on the $g$-axis indictate the various critical values, $g_{c,n}$, in the $n^{th}$ region, while double arrows indicate the $g_0^{(n)}$ 
for which the scattering length vanishes.}
\end{center}
 
 \end{figure}
From this figure, we immediately notice the similarities between the level rearrangements as we move from one region to the next, along a fixed value for the energy, $E$. 
Each region begins at $a=0$, undergoes a level rearrangement, and then returns to $a=0$.  In the language 
of the ZS, each region corresponds to one complete $2\pi$ CW rotation along the spiral. 
Figure 5 suggests that it may be possible to map every $g_n$ ($n\neq0$) region onto $g_0$ by some appropriate scaling of the $g$-axis. 
This mapping should ensure that $a=0$ in, say, $g_1$,  matches with $a=0$ in $g_0$, and that the critical $g_1$ value in the first region overlaps with the critical $g_0$ value in the zeroth-region, and so on.  

Let us first define some useful nomenclature.   We define $g_0^{(n)}$ as the $n^{th}$ value of $g$ for which $a=0$ (double arrows in Fig.~ 5).  Next, $g_{c,n}$ is defined as the $n^{th}$ critical $g$ value; that is the $g$ value at which the $n^{th}$ level rearrangement occurs (single arrows in Fig.~5). Lastly, $\tilde{g}$ is the value of $g$ outside the region $g_0$ which we intend to map back into $g_0$. 
For example, consider the point labeled by the solid dot in the $g_2$ region of Figure 5. Considering this point, which we wish to map back into $g_0$ (represented by the open circle in Fig.~5), we have 
$\tilde{g}=75$, $g_{c,2}=25\pi^2/4$ and $g_0^{(2)} \simeq 59.67951594410...$. The mapping that takes this $\tilde{g}$ back into $g_0$ is
\begin{equation}
g = \frac{(\tilde{g}-g_0^{(2)})g_{c,0}}{g_{c,2}-g_0^{(2)}}\simeq 18.84894603...,
\end{equation}   
where $g_{c,0}=\pi^2/4$ is the zeroth critical $g$ value.  We may generalize this example to  {\em any} region by employing the following prescription
\begin{equation}
g = \frac{(\tilde{g}-g_0^{(n)})g_{c,0}}{g_{c,n}-g_0^{(n)}}~,~~~~(n\neq 0).
\end{equation}
With this remapped value of $g$, we also have the associated energy, $E$.  If the mapping is indeed exact, the energy $E$ of the remapped point {\em should} be identical to the  
the energy for the same $g$ value in $g_0$.  

In Fig.~6 we study this mapping for the FSW with $b/\el = 0.01$. At first glance, the left panel Fig.~6 appears to be show that the mapping is exact, but a closer examination of the spectrum for values of 
$g$ near the resonance (right panel in Fig.~6) reveals 
noticeable discrepancies between data in the $g_0$ and 
$g_n>g_0$ regions.  Remarkably, even for $b/\el = 0.01$, the mapping of the data from $g_1$ and $g_2$ into $g_0$ agree almost perfectly ({\it i.e.,} the dashed (red online) and the dot-dashed (green online) curves, respectively).   Regardless, the mapping given by Eq.~(10) is {\em not exact} for any finite range, $b$.
\begin{figure}[h]
\begin{center}
\includegraphics[scale=.35]{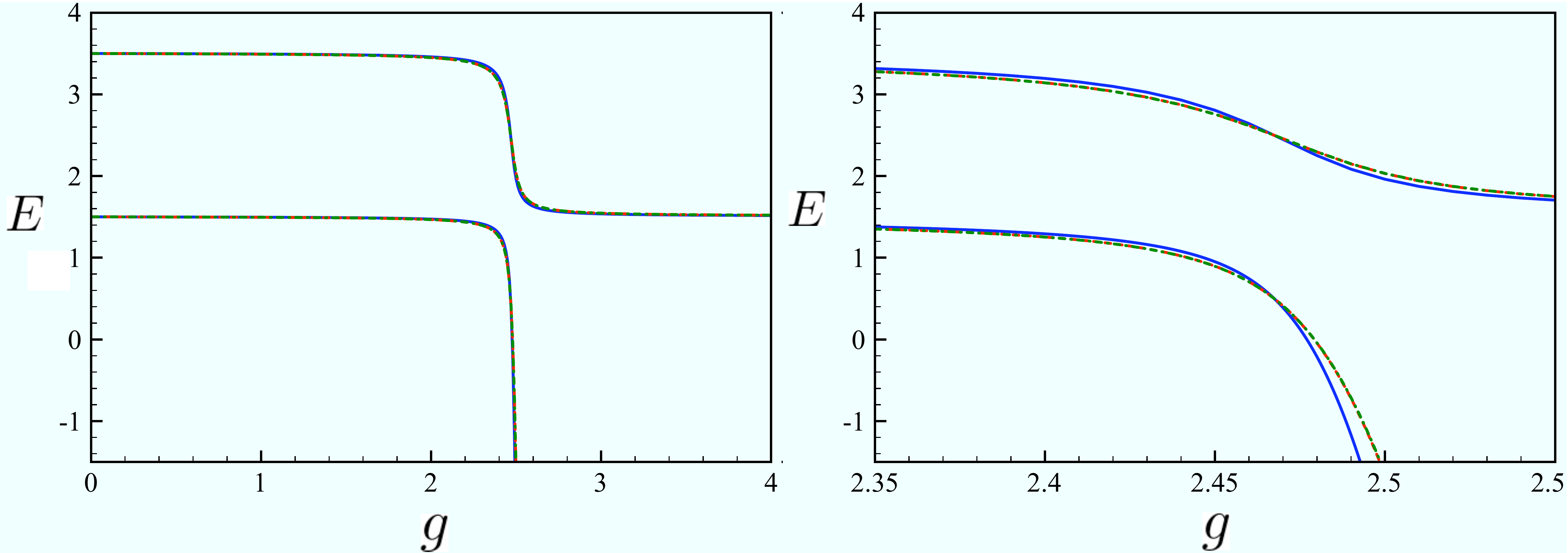}
 \caption{Left:  Equation (10) as applied to data from the FSW to different regions with $b/\el=0.01$. Right: The same data as in the left-panel, but with $g$ values surrounding the critical value, $g_{c,0}$.
In both panels, the solid (blue online), dashed (red online) and dot-dashed (green online) curves correspond to the $g_0, g_1$ and $g_2$ regions, respectively.  In these figures, the remapped
data from $g_1$ and $g_2$ are indistinguishable on the scale of the plots.}
\end{center}
\end{figure}

We can, however, show that this mapping becomes exact in the $b\to0$ limit by considering Equation (\ref{otd}). 
We write this expression in the notationally convenient form $\tilde{a}(g)=R(E)$ where $R(E)=\frac{\sqrt{2}\Gamma(1/4-E/(2\hbar\omega)}{\Gamma(3/4-E/(2\hbar\omega))}$and $\tilde{a}(g)=2a(g)/\el$. Our goal is now to map $g$ values in two different regions, $g_k$ and $g_{k'}$, onto a value in the region $g_0$ and investigate the difference in their energy values. We denote $\delta E = E_{k'}-E_k$ and note that we have the two expressions for the spectrum $\tilde{a}(g_k)=R(E_k)$ and $\tilde{a}(g_{k'})=R(E_{k'})$. The difference between these two expressions is $\tilde{a}(g_{k'})-\tilde{a}(g_k) = R(E_{k'})-R(E_{k})= R(\delta E +E_k)-R(E_k)$. Taylor expanding up to first order in $\delta E$ gives 
\begin{equation}
\delta E = \frac{\delta a}{R'(E_k)},
\end{equation}
where $\delta a  = \tilde{a}(g_{k'})-\tilde{a}(g_k)$ and $R'(E_k) = \frac{d R(E_k)}{d E_k}$.  From  Eq.~(\ref{otd}) we may re-express Eq.~(11) as
\begin{equation}
\delta E = \frac{\delta a}{\tilde{a}'(g_k)}\frac{d E_k}{d g_k},
\end{equation}      
which upon noting that $E_k = R^{-1}(a(g_k))$, becomes the implicit expression
%(from which it follows $\frac{d E_k}{d g_k} = \frac{\tilde{a}'(g_k)}{R^{-1}(R'(a(g_k)))}$)
\begin{equation}\label{LHSRHS}
R'(a(g_k)) = R\left(\frac{\delta a}{\delta E}\right).
\end{equation} 
Assuming $\delta E$ to be small compared to $\delta a$, we seek an asymptotic expression for 
 \begin{equation}
  R\left(\frac{\delta a}{\delta E}\right) =  \frac{\sqrt{2}\Gamma(1/4-\frac{\delta a}{2\delta E})}{\Gamma(3/4-\frac{\delta a}{2\delta E})}.
  \end{equation}
An application of Euler's reflection formula\cite{handbook} and Stirling's approximation to the above gives the approximate expression
  \begin{equation}
  R\left(\frac{\delta a}{\delta E}\right) \simeq   2\sqrt{\frac{\delta E}{\delta a}} \tan{\left(\frac{\pi \delta a}{2\delta E}-\frac{3\pi}{4}\right)}.
  \end{equation}
Equation (15), along with Eq. (\ref{LHSRHS}) gives
 \begin{equation}
\frac{\delta a }{\delta E} \simeq \frac{2}{\pi} \cot^{-1}\left( 2 \sqrt{\frac{\delta E}{\delta a}}\frac{1 }{R'(a(g_k))} \right)+\frac{3}{2}.
 \end{equation}
With the approximation $\cot^{-1}(x)\simeq \pi/2$, $x\ll1$ the difference in the energies becomes
 \begin{equation}
 \delta E = \frac{2}{5} \delta a,
 \end{equation}
 or, defining $a(g)=bA(g)$, 
 \begin{equation}\label{errE}
 \delta E = \frac{4b}{5\el } \left(A(g_{k'})-A(g_k)\right) .
 \end{equation}
Equation (\ref{errE}) analytically shows that $\delta E\to0$ as $b\to0$, and  the $g$ values in the two different regions get mapped back into the the $g_0$ region at the {\em exact} same energy.  Therefore,
the mapping of all subsequent regions, $g_n$ $(n\ne 0)$ back onto $g_0$ is exact in the zero-range limit.  It is important to note that our analysis has not relied upon specifying the details of 
interaction, and so is equally valid for {\em any} short-range two-body interaction supporting bound states.

It is also instructive to consider how the shape of the level rearrangements curves evolve as $b \to 0$.  The shape-dependence of the curves can be established by 
expanding the right hand side of Eq.~(\ref{otd}) about $g=g_c$ (some resonant strength value) and the left hand side about $E=E_c$ where, in 3D, $E_c= 1/2, 5/2, 9/2$,.....({\it i.e.,} the energies at the back
of the Zel'dovich spiral).   The result is
\begin{equation}
\label{expand}
\frac{c_Lb}{g/g_c-1} = \frac{  c_R\el}{1-E/E_c},
\end{equation}
where $c_L$ and $c_R$ are constants unimportant to our overall discussion. Choosing two points equally spaced away from $g_c$, call these $g_1=g_c-\Delta g$, $g_2=g_c+\Delta g$, and their corresponding energies $E_1=E_c+\Delta E$, $E_2=E_c-\Delta E$ we may use two versions of the approximation in Eq. (\ref{expand}) to write
\begin{equation}
\frac{\Delta g}{g_c} = \frac{b}{\el} \frac{c_L \Delta E}{c_R E_c}.
\end{equation}   
The important point to take away from this analysis is that the width of the rearrangement region, {\it i.e.} the range in $g$ over which the rearrangement occurs, is $\frac{\Delta g}{g_c} \sim b/\el$. An
analogous result to Eq.~(20) is briefly discussed in Ref. [\onlinecite{Kolo}] in the context of exotic atoms.  There,  the width is stated to be $\sim b/a_B$, where $a_B$ is the Bohr radius. We see that our Eq.~(20)
is consistent with the result for exotic atoms, in that the width of the rearrangement region is of the order of the range of the potential over the characteristic length of the problem.    

In Fig.~7, we numerically verify our analytical expression, {\it viz.,} Eq.~(20), by plotting the lowest two branches of the FSW for decreasing values of the range, $b$, of the potential.  It is evident that as $b \to 0$, the
level rearrangements curves evolve to a series of staircase functions, which is entirely expected given the collective results of Equations (18) and (20).
\begin{figure}[h]
\begin{center}
\includegraphics[scale=.45]{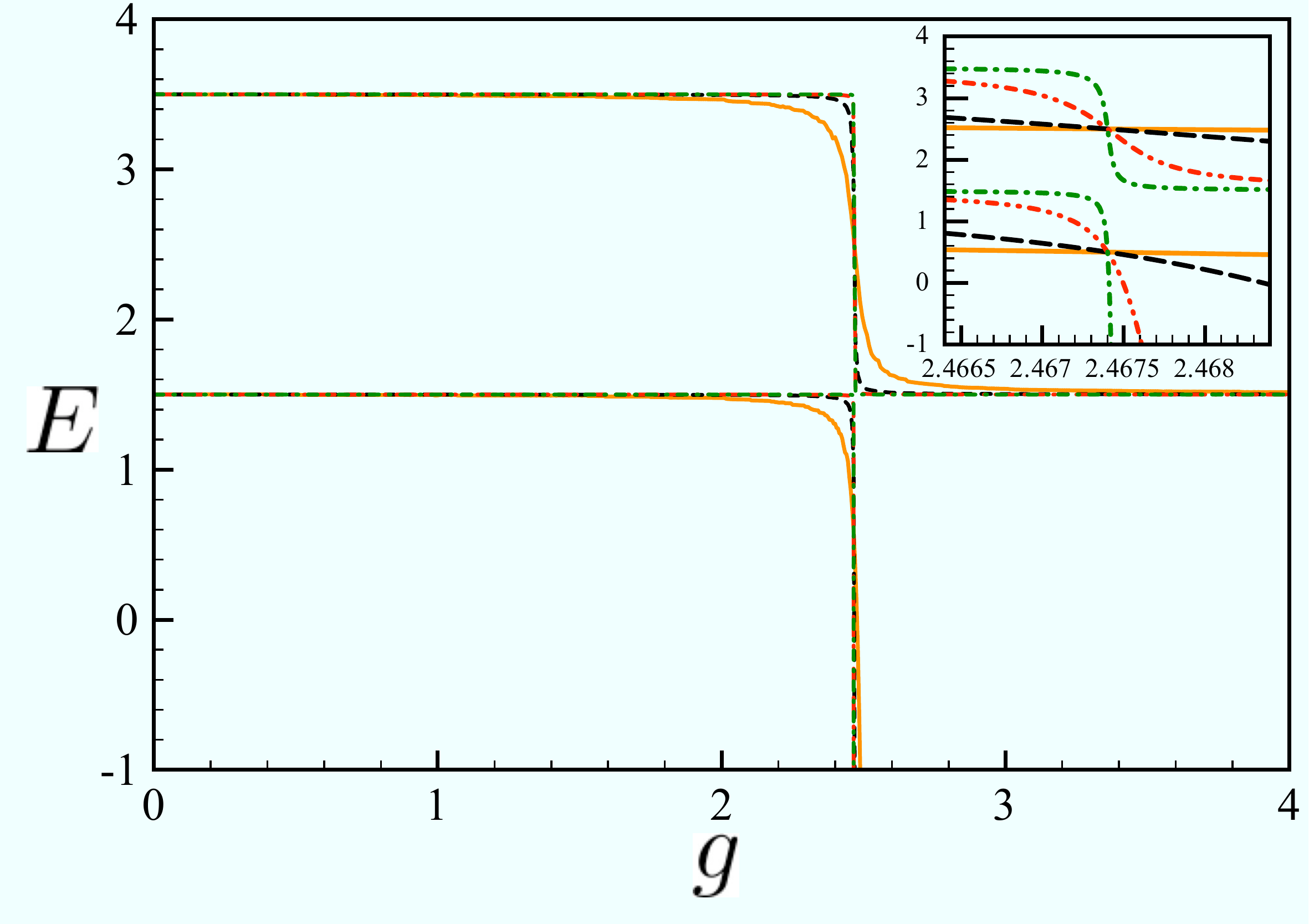}
 \caption{(Color online) Level rearrangement spectrum for several different values of the range, $b$ for the FSW. The solid line (orange online) is $b/\el=0.01$, The dashed line (black online) is $b/\el=0.001$, the 
 dot-dot-dash line (red online) $b/\el=0.0001$ and the dot-dash line (green online) is $b/\el=0.00001$. Inset: A magnification of the data in the main figure near $g_{c,0}$ further illustrating how the level 
rearrangements evolve to the staircase profile as $b \to 0$}
\end{center}
 \end{figure}
This staircase property of the spectrum has also been discussed in Refs. [\onlinecite{Kolo},\onlinecite{ostro}] in the context of the quantum defect of atomic physics, but using an entirely different approach to the one presented here.
Note that in the inset to Fig.~7, all four curves intersect at a common point, namely, at $g=g_c$ which corresponds to the back ({\it i.e.,} $|\phi| = \pi$) of the Zel'dovich spiral.

There are two noteworty points to be taken from this staircase like behaviour. The first is that any other panel of the spectrum, {\it e.g.}~$g_1$,~$g_2$ {\it etc.} in Fig.~5 (provided $b/\el \ll 1$),
can be obtained by simply applying the scaling transformation, Eq.~(10), to the data in the
$g_0$ region . In addition, the staircase property of the level rearrangements as $b \to 0$ is not specific to the FSW, which implies that for {\em any} short-range two-body potential, the $E$ vs.~$g$ curves
will exhibit the same scaling symmetry provided the critical values, $g_{c,n}$ are properly scaled as $b \to 0$.  It is also important to realize that even the staircase level rearrangements
are mapped onto the universal $E$ vs.~$a$ spectrum, just as with the other potentials listed in Eq.~(7) with $b/\el \ll 1$ in the right panel of Figure 1.

\section{Conclusions}
In this paper, we have examined the two-body problem of ultra-cold harmonically trapped interacting atoms and its relation to the Zel'dovich effect. We have shown, through our construction of the ``Zel'dovich spiral'',
that the universal spectrum in terms of the scattering length is exactly equivalent to the Zel'dovich effect.  This non-trivial observation has been used to motivate further experimental studies in order to provide additional data for the
$E$ vs.~$a$ spectrum, which may then be used to establish the first direct experimental obseravtion of the Zel'dovich effect.  
Finally, we have shown that in the $b \to 0$ limit, the level rearrangement spectrum exhibits
an exact scaling symmetry, which has until now, gone unnoticed.  The exact mapping means that the {\em entire} $E$ vs.~$g$ spectrum (and therefore the $E$ vs.~$a$ spectrum) 
may be obtained solely from  knowledge of the $g_0$ region as $b \to 0$.

\section{acknowledgements}
Z. MacDonald would like to acknowledge the Natural Sciences and Engineering Research Council of Canada
(NSERC) USRA program for financial support.  B. P.  van Zyl and A. Farrell would also like to acknowledge the NSERC Discovery Grant program for additional
financial support.


\begin{thebibliography}{99}

\bibitem{Zel} Y. B. Zel'dovich, Sov. J. Solid State, {\bf 1}, 1497 (1960).
\bibitem{comb1} M. Combescure, A. Khare, A. Raina, J.M. Richard and C. Weydert, Int. J. Mod. Phys. B {\bf 21} 3765 (2007).
\bibitem{comb2}  M. Combescure, C. Fayard, A. Khare,  and J. M. Richard, J.Phys. A: Math. Theor. {\bf 44} 275302 (2011).
\bibitem{Kolo} E.B. Kolomeisky and M. Timmins, Phys. Rev. A, {\bf 72}, 022721 (2005).

\bibitem{karnakov}B. M. Karnakov and V. S. Popov, JETP {\bf 97}, 890 (2003).
\bibitem{cohen} The reader
will find a clear description of the underlying physics of Feshbach
resonance in Cohen-Tannoudji's lecture-notes~ 
``Atom-atom interactions in 
 ultra-cold quantum gases'', in {\it Lectures on Quantum Gases}, 
Institut Henri Poincar\'e, Paris, April 2007.
\bibitem{stoferle}  T. St\"oferle {\it et al.} Phys. Rev. Lett. {\bf 96}, 030401 (2006).
\bibitem{farrell2} A. Farrell and B. P. van Zyl,  J. Phys. A: Math. Theor. {\bf 43}, 015302 (2010).
\bibitem{busch} T. Busch, B-G Englert, K. Rzazewski and M. Wilkens, Foundations of Physics, {\bf  28}, 549 (1998).

\bibitem{shea2008} P. Shea, B. P. van Zyl and R. K. Bhaduri, Am. J. Phys. {\bf 77}, 511 (2008).
\bibitem{handbook}
Abramowitz and Stegun, {\it Handbook of Mathematical Functions} (Dover, New York, 1970).
\bibitem{poschl} G. P\"oschl and E. Teller,  Z. Phys., {\bf 83}, 143 (1933).
\bibitem{exp} J. Shapiro and M. A. Preston, Can. J. Phys. {\bf 34},  451 (1956).


\bibitem{Flugge} S. Flugge, Practical Quantum Mechanics, Springer-Verlag, Berlin-Heidelberg-New York, 1971.
\bibitem{ahmed} Z. Ahmed, Am. J. Phys. {\bf 78} 418, (2010). 
\bibitem{CJP} A. Farrell, B.P. van Zyl, Can. J. Phys. {\bf 88}, 817 (2010).
\bibitem{ostro} V. N. Ostrovsky, Phys. Rev. A {\bf 74} 012714 (2006).






%\bibitem{grein} M. Greiner, Mandel, T. Esslinger, T.W H\"ansch and I. Bloch, Nature London {\bf 415}, 39 (2002) 
%\bibitem{fesh} H. Feshbach, Ann. Phys. (N.Y.) {\bf 281}, 519 (2000)
%\bibitem{kudr}A. E. Kudryavtsev, V.E. Markushin and I.S. Shapiro, Zh. Eksp. Teor. Fiz. {\bf 74}, 432-444 (1978)
%\bibitem{shapiro} I.S. Shapiro, Phys. Rept. {\bf 35}, 129 (1978). 
%\bibitem{Gal} A. Gal, E. Friedman and C.J. Batty, Nuclear Physics A, {\bf 606}, 283 (1996).
%\bibitem{Karn} B. M. Karnakov, and V. S. Popov, JETP 97, 890Ð914 (2003).



%\bibitem{note1} The mapping is one-to-one everywhere but at $\pm \infty$, where mathematicians would prefer to talk about branch cuts.  These
%mathematical details are not essential to the present paper, and in our opinion rather serve to distract attention away from the utility of the mapping.




%%%%%
\end{thebibliography}
\end{document}